\documentclass[10pt,aps,prd,amsmath,amssymb,nofootinbib,superscriptaddress,preprintnumbers,nobibnotes,floatfix,notitlepage]{revtex4-1}

\usepackage{graphicx}
\usepackage{dcolumn}
\usepackage{bm}
\usepackage{cancel}
\usepackage{physics}
\usepackage[dvipsnames]{xcolor}
\usepackage{booktabs}
\usepackage{multirow}
\usepackage{tensor}
\usepackage{fancyhdr}
\usepackage{hyperref}
\usepackage[normalem]{ulem}
\usepackage{mathtools}
\usepackage{tikz}
\usepackage{enumitem}
\usetikzlibrary{calc}
\usepackage{graphicx}
\usepackage{stackrel}
\usepackage[caption=false]{subfig}

\usepackage{empheq}

\usepackage{tikz}
\usetikzlibrary{decorations.pathreplacing, arrows.meta}

\def\rh{r_{h}}
\def\rc{r_{c}}

\newcommand{\bea}{\begin{eqnarray}}
\newcommand{\eea}{\end{eqnarray}}

\newcommand{\be}{\begin{equation}}
\newcommand{\ee}{\end{equation}}
\newcommand{\lf}{\left}
\newcommand{\rg}{\right}
\newcommand{\pa}{\partial}

\numberwithin{equation}{section}

\begin{document}

\preprint{CCTP-2025-12, ITCP-IPP 2025/12}

\title{Scalar field scattering in a Schwarzschild-de Sitter geometry}

\author{Marco de Cesare}
\email{marco.decesare@na.infn.it}
\affiliation{Scuola Superiore Meridionale, Largo S. Marcellino, 10, 80138 Napoli, Italy}
\affiliation{INFN sezione di Napoli, via Cintia, 80126 Napoli, Italy}

\author{Marcello Miranda}
\email{m.miranda@ssmeridionale.it}
\affiliation{Scuola Superiore Meridionale, Largo S. Marcellino, 10, 80138 Napoli, Italy}
\affiliation{INFN sezione di Napoli, via Cintia, 80126 Napoli, Italy}

\author{Achilleas P. Porfyriadis}
\email{porfyriadis@physics.uoc.gr}
\affiliation{Crete Center for Theoretical Physics, Institute of Theoretical and Computational Physics,\\
Department of Physics, University of Crete, 70013 Heraklion, Greece}

\begin{abstract}

We solve analytically the low-frequency s-wave dynamics of a massless scalar field propagating on a Schwarzschild-de Sitter black hole background. A rigorous application of the method of matched asymptotic expansions allows us to connect the scalar's evolution in the proximity of the black-hole horizon with that on cosmological scales. The scattering coefficients, greybody factors, and Wigner time delay are computed explicitly. 
We consider both small and large black holes, with black-hole to cosmological horizon radii parametrically small and of order unity, respectively. This extends previous studies confined to the small black-hole regime only. In addition, for small black holes we perform a calculation that remains agnostic about the relative size between the ratio of the geometry's horizons and the scalar's frequency in units of the black-hole radius.
When the two are comparable, we find that they are interchangeable in the greybody factor, which is symmetric under $\omega\leftrightarrow 1/r_c$ (where $\omega$ is the scalar's frequency and $r_c$ the cosmological horizon radius).

\end{abstract}

\maketitle

\nopagebreak

\section{Introduction}

The scattering of minimally coupled massless matter fields off a spherically symmetric and asymptotically flat black hole displays well-known universality properties in the low-frequency limit \cite{Das:1996we}. 
However, these universalities do not extend to non-asymptotically flat spacetimes. In particular, in the asymptotically de Sitter case, there exists evidence \cite{Kanti:2005ja, Harmark:2007jy} that the greybody factor of a small Schwarzschild-de Sitter black hole is non-vanishing in the zero frequency limit, and is given by a simple function of the black-hole and cosmological horizon radii. The situation for a general large black hole, with a finite ratio of black-hole to cosmological horizon, is unknown.
On the other hand, for non-minimally coupled scalars the greybody factor can be vanishing at zero frequency \cite{Crispino:2013pya, Kanti:2014dxa}.

Schwarzschild-de Sitter is an exact spherically symmetric solution of vacuum general relativity with a positive cosmological constant. As such, it provides the simplest model of a black hole embedded in an expanding cosmology dominated by dark energy, and serves as a testbed for understanding the imprints of cosmological dynamics on localized systems. 
Previous studies have highlighted the impact of a positive cosmological constant on black-hole collisions \cite{Zilhao:2012bb}, the dynamics of thick accretion disks \cite{Rezzolla:2003re}, and gravitational collapse \cite{Shibata:1993fx,Markovic:1999di,Musco:2004ak}.
Perturbative calculations on a Schwarzschild-de Sitter background allow us to analytically explore the interplay between the dynamics in the proximity of the black-hole horizon and the dynamics in the asymptotic de Sitter region. To leading order in the perturbative expansion, matter fields can be treated in a test-field approximation, thus neglecting backreaction on the background geometry. In this work, we consider the scattering of a massless scalar field in the low-frequency limit. In this case, the spherical wavelet component gives the dominant contribution to the transmission from the asymptotic region to the black-hole horizon, whereas higher multipole components are suppressed due to the centrifugal potential barrier. For the same reason, a massless scalar field also dominates compared to other matter fields with non-zero spin.

The paper is organized as follows. 
In Section~\ref{Sec:Dynamics}, we recast the scalar s-wave equation in a Schwarzschild-de Sitter geometry in dimensionless form, and identify the relevant physical parameters that govern the dynamics: the ratio of the event horizon to cosmological horizon radii, and the product of the radiation frequency with the black-hole horizon radius. We identify two physically relevant regimes, corresponding to small and large black holes (with respect to the cosmological horizon). These regimes are studied in the low-frequency limit, using a rigorous application of the method of matched asymptotic expansions, in Sections~\ref {Sec:regime1} and~\ref {Sec:regime2}, respectively. In both cases, analytical solutions are obtained for the scalar field, which we use to compute the scattering coefficients and observables such as the greybody factor and the Wigner time delay. 
Section~\ref{Sec:conclusions} further discusses our main results and methodological merits.

\section{Massless Scalar Field in Schwarzschild-de Sitter}\label{Sec:Dynamics}
The line element of the Schwarzschild-de Sitter spacetime is
\be
    \dd s^2=-f(r) \dd t^2+f(r)^{-1}\dd r^2+r^2\dd\Omega^2~,
\ee
where
\be\label{Eq:metric}
    f(r)=1-\frac{2M}{r}-\frac{\Lambda}{3}r^2=-\frac{\Lambda}{3r}(r-\rh)(r-\rc)(r+\rc+\rh)~.
\ee
Here $M$ is the black-hole mass and $\Lambda$ the cosmological constant, while $\rh$ and $\rc$ denote the black-hole and cosmological horizon radii, respectively. The two sets of parameters are related by
\be
    M=\frac{\rc \rh (\rc+\rh)}{2 \left(\rc^2+\rc \rh+\rh^2\right)}~,\qquad \Lambda=\frac{3}{\rc^2+\rc \rh+\rh^2}~.
\ee
The case of an asymptotically flat Schwarzschild black hole corresponds to the $r_c\to+\infty$ limit.

We study the dynamics of a massless scalar field in the test field approximation on such a background, which is governed by the scalar wave equation
\be\label{Eq:KGcovform}
    \Box\phi=0~.   
\ee
Assuming the scalar field has no dependence on the angular coordinates, $\phi=\phi(t,r)$\,, Eq.~\eqref{Eq:KGcovform} becomes
\be\label{Eq:KG_tr_coords}
\begin{split}
&-\frac{r \left(r_h^2+r_h r_c+r_c^2\right)}{(r_c-r)(r-r_h)(r+r_h+r_c)}\,\pa_t^2\phi(t,r)+\frac{(r_c-r)(r-r_h) (r+r_h+r_c)}{r \lf(r_h^2+r_h r_c +r_c^2\rg)}\, \pa_r^2\phi(t,r)\\
&-\frac{4 r^3-2 r \lf(r_h^2+r_h r_c+r_c^2\rg)+r_h r_c (r_h+r_c)}{r^2 \lf(r_h^2+r_h r_c+r_c^2\right)} \,\pa_r\phi(t,r)=0~.
\end{split}
\ee
We will work in the frequency domain, expressing the scalar field as a Fourier integral $\phi(t,r)=\int_{-\infty}^{\infty}\frac{\dd \omega}{2\pi} e^{-\text{i$\omega $} t} \phi_\omega(r)$\,. 
We also find it convenient to describe the region between the two horizons, $r_h<r<r_c$\,, in terms of a new dimensionless radial coordinate $x\in(0,\infty)$ given by 
\be\label{Eq:def_xofr}
x(r)=\frac{r_c}{r_h}\frac{r-r_h}{r_c-r}~ \quad \Longleftrightarrow \quad  r(x)= \frac{r_h r_c(1+x)}{r_c +r_h\,x}~,
\ee
with $x(r\to r_h)=0$ and $x(r\to r_c)=\infty$\,.
Notice that the above change of coordinate is well defined only if $r_h\neq0$ and $r_h\neq r_c$\,.

In this work we will analyze the dynamics in the low-frequency limit using the method of matched asymptotic expansions. To this end, it is beneficial to work with a rescaled scalar field variable $\psi_\omega(x)= x\,\phi_\omega(x)$, in terms of which Eq.~\eqref{Eq:KG_tr_coords}, in frequency domain, reads:
\be\label{Eq:KG_xs_coord}
\begin{split}
x^2 (1 + x) \, \psi_\omega''(x)
-\frac{1+2\alpha+3\alpha(1+2\alpha)x+\alpha^2 (7+2\alpha)x^2+\alpha^2(2+\alpha)x^3}{(1 + \alpha x)\left[1 + 2\alpha  + \alpha (2 + \alpha)x \right]}\,x\, \psi_\omega'(x)&\\[5pt]
+\left(\frac{1+2\alpha+3\alpha(1+2\alpha)x+\alpha^2 (7+2\alpha)x^2+\alpha^2(2+\alpha)x^3}{(1 + \alpha x)\left[1 + 2\alpha  + \alpha (2 + \alpha)x \right]}+\frac{ \epsilon^2(1+\alpha+\alpha^2)^2 (1 + x)^3 }{(1-\alpha)^2\left[1 + 2\alpha  +\alpha (2 + \alpha)x\right]^2}\, \right)\psi_\omega(x)& =0~,
\end{split}
\ee
where a prime denotes derivative with respect to $x$\,, and we have introduced the dimensionless parameters $\alpha=\rh/\rc$\,, and $\epsilon=\omega\,\rh$\,. In the following, we will assume, without loss of generality, that $\epsilon>0$ and the low-frequency limit is:
\[\epsilon\ll1\,.\]
We observe that the $\epsilon\to0$ limit is a singular one for the dynamics, as the behaviour of solutions is dramatically different for $\epsilon\neq0$ and $\epsilon=0$\,. In particular, when $\epsilon$ is nonzero the scalar field is freely oscillating in the proximity of both horizons, whereas for $\epsilon=0$ it has a divergent behaviour (consistently with no-hair theorems for minimally coupled massless scalars \cite{Bekenstein:1995un,Torii:1999,Winstanley:2002jt}). The $\alpha\to0$ limit corresponds to pushing the cosmological horizon to an infinite distance from the black hole. This too is a singular limit for the dynamics, known to be associated e.g.~with a spectral instability \cite{Zhou:2025xta}.

Equation~\eqref{Eq:KG_xs_coord} can be solved analytically using matched asymptotic expansions, which is the method of choice for studying singular perturbation problems. In this paper we focus on the following two regimes:
\begin{enumerate}[label=\arabic*)]
\item Small black holes, with $r_h\ll r_c$\,.
In this regime, $\alpha\ll 1$ and we will remain agnostic with regard to the relative size of $\alpha$ and $\epsilon$\,. This means, in particular, that our analysis includes both sub-horizon modes with $\omega r_c\ll 1$ and super-horizon modes with $\omega r_c\gg 1$\,;
\item Large non-extremal black holes, with $r_h$ and $r_c$ comparable. In this regime, $\alpha$ and $(1-\alpha)$ remain both finite.
\end{enumerate}
A third regime of interest, not studied in this work, is the regime corresponding to near-extremal (Nariai) black holes with $(1-\alpha)\ll 1$\,. In this case, near the extremal horizon a $dS_2\times S^2$ scaling geometry opens up and warrants separate investigation \cite{Nariai:1999reprint,Ginsparg:1982rs,Cardoso:2004uz,Katona:2023vtq}.

\section{Small black holes}\label{Sec:regime1}
In this regime, both physical parameters governing the dynamics of \eqref{Eq:KG_xs_coord} are small, that is $\epsilon\ll1$ and $\alpha\ll1$\,. We partition the spacetime between the two horizons, $r_h$ and $r_c$\,, into two overlapping regions:
\begin{itemize}
\item \textit{near region}: $ x\ll 1/\max\{\alpha,\epsilon\}$~,
\item \textit{far region}: $x\gg1$~.
\end{itemize}
Given that $\max\{\alpha,\epsilon\}\ll 1$\,, the near and far regions overlap in the region $1\ll x\ll 1/\max\{\alpha,\epsilon\}$\,.

\subsection{\texorpdfstring{Near region: $x\ll 1/\max\{\alpha,\epsilon\}$}{Near region}}\label{Sec:regime1_near}

In this region, Eq.~\eqref{Eq:KG_xs_coord} can be approximated as
\be\label{Eq:sub-near-region}
 x^2 (1+x) \,\psi_\omega''(x)-x\,\psi_\omega'(x)+\left(1+\epsilon^2 \right)\psi_\omega(x)=0~\,.
\ee
Note that, as $\alpha$ does not feature in Eq.~\eqref{Eq:sub-near-region}, this equation is identical to the near-region dynamics in the Schwarzschild geometry. Equation~\eqref{Eq:sub-near-region} can be solved exactly, with general solution
\be\label{Eq:sub-near-region-solution}
\begin{split}
    \psi^{\rm(near)}_{\omega}(x)=&\,
C_{1}^{\rm(near)}x^{1-i \epsilon}
  {}_2F{}_1\left( 1-i\epsilon,\, -i\epsilon,\, 1 - 2i\epsilon;\, -x\right)+
C_{2}^{\rm(near)}x^{1+i \epsilon}
 {}_2F{}_1\left(1+i\epsilon ,\, i\epsilon ,\, 1 + 2i\epsilon;\,-x\right)~,
\end{split}
\ee
where $C_{1,2}^{\rm(near)}$ are integration constants and~${}_2F_1$ is the hypergeometric function.
This solution exhibits the following asymptotic behaviours:
\begin{subequations}\label{Eq:Regime1_Near_Asymptotics}
\begin{empheq}[left={\psi_{\omega}^{\rm(near)}(x)\approx\empheqlbrace}]{align}
\displaystyle
&\,C_{1}^{\rm(near)}\,x^{1-i\epsilon}+C_{2}^{\rm(near)}\,x^{1+i\epsilon}~, &&\text{as }x\to 0~,
\label{Eq:Regime1_Near_Asymptotics_a} \\[10pt]
\displaystyle
&\,\left(C_{1}^{\rm(near)}+C_{2}^{\rm(near)}\right)x+i\epsilon\left(C_{1}^{\rm(near)}-C_{2}^{\rm(near)}\right)~, &&\text{as } x \to +\infty~.
\label{Eq:Regime1_Near_Asymptotics_b}
\end{empheq}
\end{subequations}
Notice that, while we will perform matched asymptotic expansions to leading order only, we have kept the sub-leading $\epsilon^2$ term in the coefficient of $\psi_\omega$ in Eq.~\eqref{Eq:sub-near-region}. The reason is that this term is necessary in order to produce a solution which satisfies the correct boundary conditions at the black-hole horizon: regularity at the horizon requires a solution which behaves like $\phi_\omega\sim e^{\mp i\omega r_*}$, where $r_*=\int \frac{dr}{f(r)}$ is the tortoise coordinate, and this matches precisely the $x^{1\mp i\epsilon }$ solutions in Eq.~\eqref{Eq:Regime1_Near_Asymptotics_a}.

\subsection{ \texorpdfstring{Far region: $x\gg1$}{Far region}}
At large distances from the black-hole horizon, we can approximate Eq.~\eqref{Eq:KG_xs_coord} as
\be\label{Eq:sub-far-region}
x^4 \, \psi_{\omega}''(x)
-\frac{2 \alpha^2 x^5 }{(1 + \alpha x)(1 + 2 \alpha x)}\, \psi_{\omega}'(x)+    
\left(\frac{2 \alpha^2 x^4 }{(1 + \alpha x)(1 + 2 \alpha x)}+\frac{\epsilon^2 x^4 }{(1 + 2 \alpha x)^2}\right) \psi_{\omega}(x)=0~.
\ee
The exact solution to this far region equation is
\be\label{Eq:sub-far-region-solu}
\begin{split}
    \psi^{\rm (far)}_{\omega}(x)=C_{1}^{\rm (far)}\,(1 + 2 \alpha x)^{- \frac{i \epsilon}{2\alpha}}{ \left( \alpha^2 x + i \alpha\epsilon x +i \epsilon \right) }+C_{2}^{\rm (far)}\,(1 + 2 \alpha x)^{\frac{i \epsilon}{2\alpha}}{ \left( \alpha^2 x - i \alpha\epsilon x - i \epsilon \right)}~,
\end{split}
\ee
where $C_{1,2}^{\rm (far)}$ are integration constants. The above solution has the following asymptotics:
\begin{subequations}\label{Eq:Regime1_Far_Asymptotics}
\begin{empheq}[left={\psi_{\omega}^{\rm(far)}(x)\approx\empheqlbrace}]{align}
\displaystyle
&\,\left[C_{1}^{\rm (far)} \left(\alpha ^2+i \alpha  \epsilon +\epsilon ^2\right)+C_{2}^{\rm (far)} \left(\alpha ^2-i \alpha  \epsilon +\epsilon ^2\right)\right]x+i \epsilon\left( C_{1}^{\rm (far)}  - C_{2}^{(far)} \right)~, && \text{as } x \to 0~,
\label{Eq:Regime1_Far_Asymptotics_a} \\[10pt]
\displaystyle
&\,C_{1}^{\rm (far)}  (\alpha^2 +i\alpha \epsilon) (2\alpha  x)^{-\frac{i \epsilon }{2 \alpha }} \, x+  C_{2}^{\rm (far)} (\alpha^2 -i \alpha\epsilon) (2\alpha  x)^{\frac{i \epsilon }{2 \alpha }}\, x~, && \text{as } x \to +\infty~.
\label{Eq:Regime1_Far_Asymptotics_b}
\end{empheq}
\end{subequations}
It is convenient to define the following auxiliary quantities:
\begin{equation}
    C_{1}^{\rm(cosmo)} \equiv  C_{1}^{\rm(far)}\, (2\alpha)^{-\frac{i \epsilon }{2 \alpha }}  (\alpha^2 +i \alpha\epsilon )~,\quad \mbox{and}\quad C_{2}^{\rm(cosmo)} \equiv C_{2}^{\rm(far)}\, (2\alpha)^{\frac{i \epsilon }{2 \alpha }}  (\alpha^2 -i \alpha\epsilon )\,.
\end{equation}
Then, the asymptotics in the proximity of the cosmological horizon can be expressed more compactly as
\be\label{Eq:sub-far-region-solu_cosmoasymtpotics}
\psi^{\rm (far)}_{\omega}(x)\approx  C_{1}^{\rm (cosmo)}  x^{1-\frac{i \epsilon }{2 \alpha }}+  C_{2}^{\rm (cosmo)} x^{1+\frac{i \epsilon }{2 \alpha }}~, \quad \mbox{as }  x\to+\infty~.
\ee

It is worth comparing \eqref{Eq:sub-far-region} with the dynamics in the `far' asymptotics in the Schwarzschild case
$\psi_{\omega}''(x)+ \epsilon^2\psi_{\omega}(x)=0$
with solutions $\psi_{\omega}(x)\sim e^{\pm i\epsilon x}$\,. The comparison with the `far' solutions in  Schwarzschild-de Sitter \eqref{Eq:sub-far-region-solu} thus shows that the $\alpha\to0$ limit is not a smooth one. In particular, we see that deviations from the dynamics in an asymptotically flat background become most relevant on scales $x\gtrsim 1/\alpha$\,, where the solution \eqref{Eq:sub-far-region-solu} transitions from a behaviour that is similar to the asymptotically flat case, to the cosmological asymptotics described by \eqref{Eq:sub-far-region-solu_cosmoasymtpotics}.

\subsection{Matching formulae and scattering coefficients}\label{Sec:Regime1_Matching}

\begin{figure}
    \centering
\subfloat[]{\scalebox{0.36}{\includegraphics{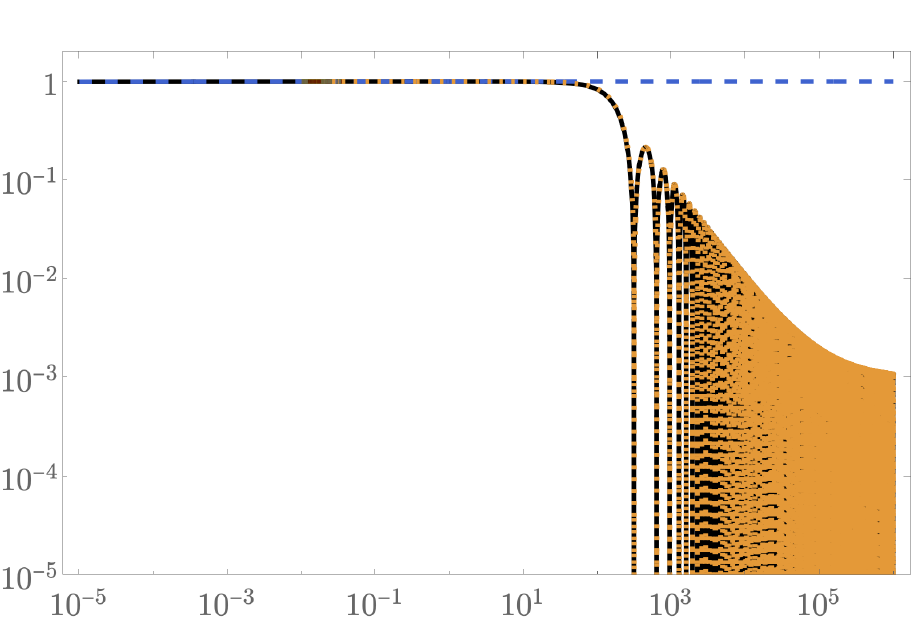}}}
\quad
\subfloat[]{\scalebox{0.365}{\includegraphics{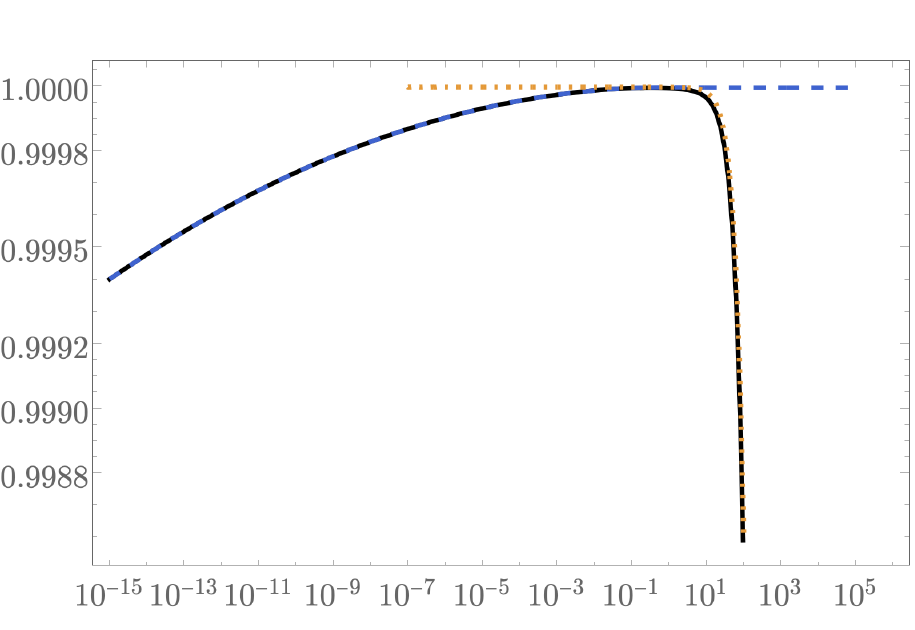}}}
\quad
\subfloat[]{\scalebox{0.368}{\includegraphics{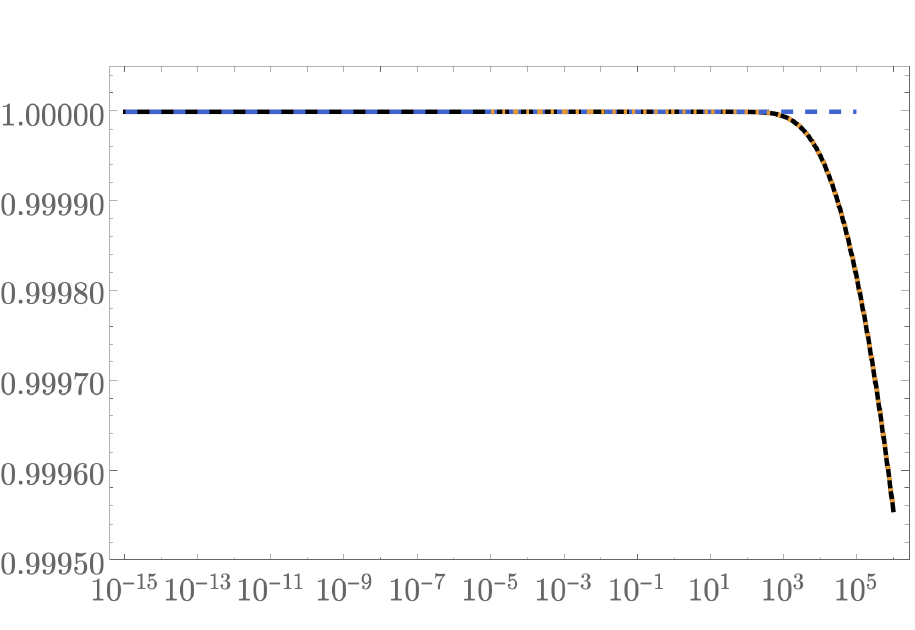}}}\\
\vspace{1mm}
\subfloat[]{\scalebox{0.365}{\includegraphics{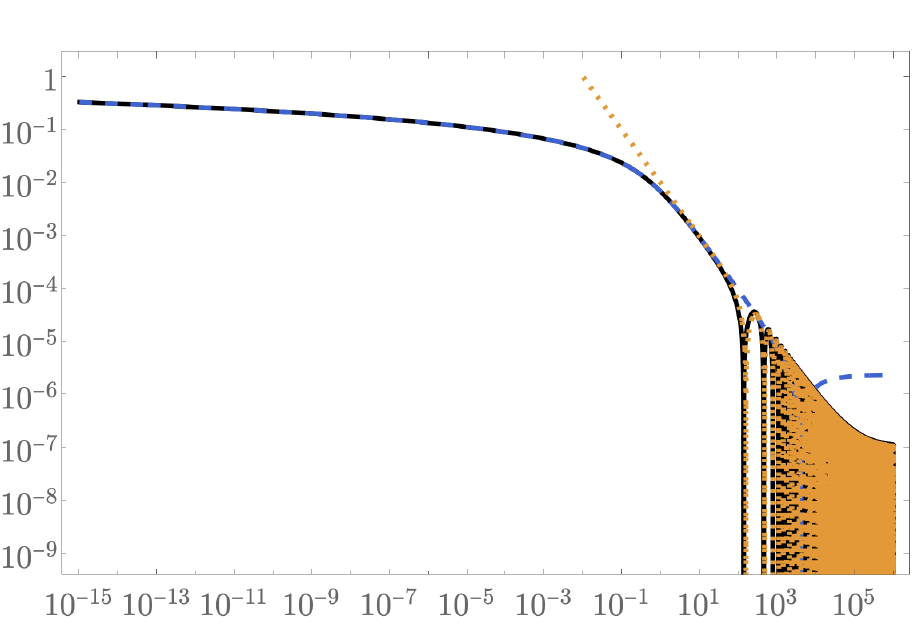}}}
\quad
\subfloat[]{\scalebox{0.365}{\includegraphics{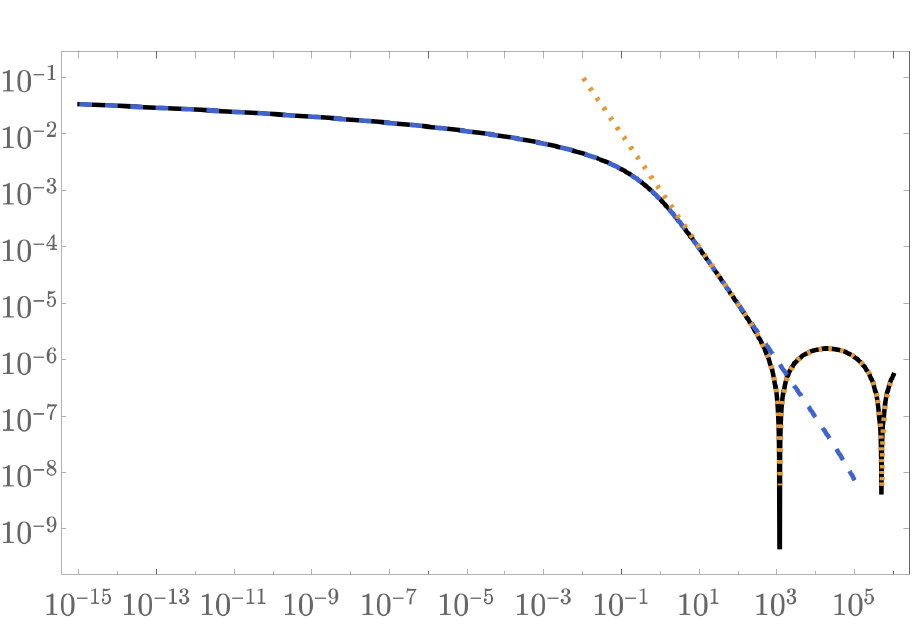}}}
\quad
\subfloat[]{\scalebox{0.365}{\includegraphics{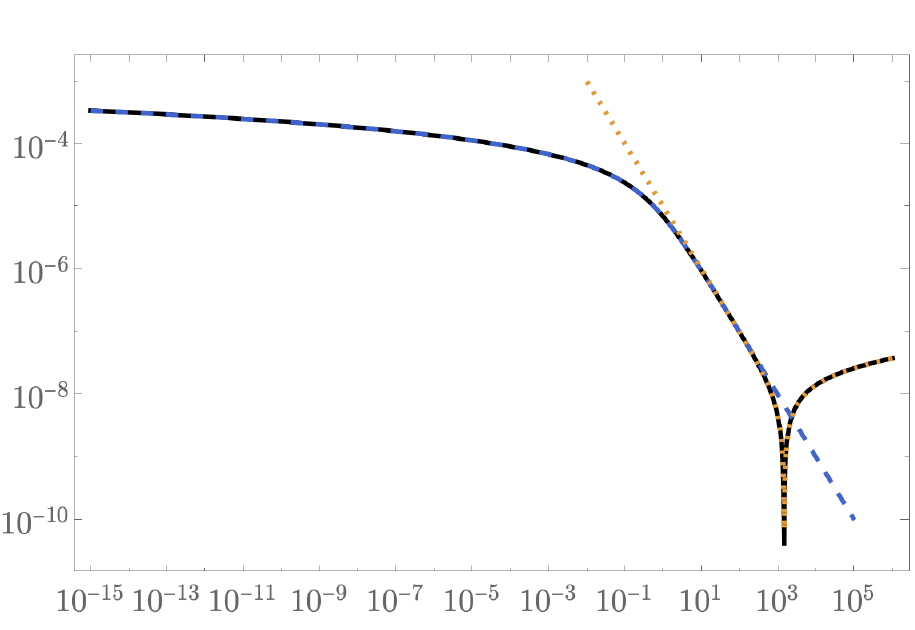}}}
\caption{The plots show the agreement between our `near' and `far' analytical solutions and a numerical one for $\phi(x)$ in the small black-hole regime. From left to right, we have  $\alpha\ll\epsilon$\,, $\alpha\sim\epsilon$\,, $\epsilon\ll\alpha$\,. Specifically, panels (a) and (d) correspond to $(\epsilon,\alpha)=(10^{-2},10^{-5})$\,, panels (b) and (e) to $(\epsilon,\alpha)=(10^{-3},10^{-3})$\,, (c) and (f) to $(\epsilon,\alpha)=(10^{-5},10^{-3})$\,. The top and bottom rows, respectively, show the real and imaginary parts of $\phi$. The thick black curve represents the numerical solution, the dashed blue curve the near-region asymptotics, and the dotted orange curve corresponds to the far-region asymptotics.
}\label{Fig:Regime1}
\end{figure}

Requiring that the near and far solutions agree in their overlap region, $1\ll x\ll 1/\max\{\alpha,\epsilon\}$\,, that is to say, matching equations~\eqref{Eq:Regime1_Near_Asymptotics_b} and~\eqref{Eq:Regime1_Far_Asymptotics_a}, we obtain the following algebraic system
\be\label{Eq:MatchingConditions_Regime1}
\begin{split}
C_{1}^{\rm(near)}+C_{2}^{\rm(near)}&=C_{1}^{\rm(far)} \left(\alpha ^2+i \alpha  \epsilon +\epsilon ^2\right)+C_{2}^{\rm(far)} \left(\alpha ^2-i \alpha  \epsilon +\epsilon ^2\right)~,\\
\qquad C_{1}^{\rm(near)}-C_{2}^{\rm(near)}&=C_{1}^{\rm(far)}-C_{2}^{\rm(far)}~.
\end{split}
\ee
These relations constitute the unique solution to the problem of connecting the scalar wave's amplitudes at the black-hole and cosmological horizons.
For example, if we assume purely ingoing boundary conditions at the black-hole horizon, $C_2^{\rm (near)}=0$\,, then the system~\eqref{Eq:MatchingConditions_Regime1} admits a unique solution for the amplitudes of the transmitted and reflected waves, $C_{1}^{\rm(near)}$, $C_{2}^{\rm(far)}$, in terms of the amplitude of the incident wave $C_{1}^{\rm(far)}$. 
Fig.~\ref{Fig:Regime1} shows our near and far region analytical solutions in comparison to a numerical solution of Eq.~\eqref{Eq:KG_xs_coord} everywhere.

The transmission and reflection coefficients read
\begin{subequations}\label{Eq:ScatteringCoefficients}
\begin{align}
{\cal T}_\omega&\equiv \frac{r_h C_1^{\rm(near)}}{r_c C_1^{\rm(cosmo)}}=2 (2\alpha)^{\frac{i\epsilon}{2\alpha}}\frac{\alpha-i \epsilon }{1+\alpha^2+\epsilon^2-i\alpha\epsilon}~,\label{Eq:TransmissionCoefficient}\\
{\cal R}_\omega&\equiv \frac{r_c C_2^{\rm(cosmo)}}{r_c C_1^{\rm(cosmo)}}=(2\alpha)^{\frac{i\epsilon}{\alpha}}\lf(\frac{\alpha-i\epsilon}{\alpha+i\epsilon}\rg)\lf(\frac{1-\alpha^2-\epsilon^2-i\alpha\epsilon}{1+\alpha^2+\epsilon^2-i\alpha\epsilon}\rg)~.\label{Eq:ReflectionCoefficient}
\end{align}
\end{subequations}
Squaring these quantities, we obtain
\begin{subequations}
\begin{align}
|{\cal T}_\omega|^2=\frac{4(\alpha^2+\epsilon^2)}{\alpha^2\epsilon^2+(1+\alpha^2+\epsilon^2)^2}~,\label{Eq:TransmissionCoefficientSquared}\\
|{\cal R}_\omega|^2=\frac{\alpha^2\epsilon^2+(1-\alpha^2-\epsilon^2)^2}{\alpha^2\epsilon^2+(1+\alpha^2+\epsilon^2)^2}~.\label{Eq:ReflectionCoefficientSquared}
\end{align}
\end{subequations}
We observe that $|{\cal R}_\omega|^2+|{\cal T}_\omega|^2=1$ holds as an exact relation. The greybody factor is defined as $\Gamma(\omega)=|{\cal T}_\omega|^2$. Note the symmetry of $\Gamma(\omega)$ under the exchange $\alpha\leftrightarrow\epsilon$ (that is, $\omega\leftrightarrow 1/r_c$).

The results above for the scattering coefficients should be contrasted with their counterpart for a Schwarzschild background, ${\cal T}_\omega^{\rm Schw.}=-2i\epsilon/(1+\epsilon^2)$~, ${\cal R}_\omega^{\rm Schw.}=-(1-\epsilon^2)/(1+\epsilon^2)$~. We observe that the $\alpha\to0$ limit of the scattering coefficients \eqref{Eq:ScatteringCoefficients} is not smooth, due to the presence of phase factors $\sim \alpha^{\frac{i\epsilon}{\alpha}}$ which are non-analytic at $\alpha=0$\,. Nonetheless, such phase factors are absent from the greybody factor, which is indeed smooth in the $\alpha\to0$ limit, where one recovers the well-known result valid for asymptotically flat black holes in the low-frequency limit, $\lim_{\alpha\to0}\Gamma(\omega)=4\epsilon^2/(1+\epsilon^2)^4\approx 4\epsilon^2+{\cal O}(\epsilon)^4$ \cite{Das:1996we}. On the other hand, taking the $\epsilon\to0$ limit while keeping $\alpha$ fixed, we obtain
\be\label{Eq:LowFrequencyLimit1}
\lim_{\epsilon\to0}\Gamma(\omega)=\frac{4\alpha^2}{(1+\alpha^2)^2}=\frac{4r_h^2 r_c^2}{(r_h^2+r_c^2)^2}~,
\ee
which is consistent with the results found in Refs.~\cite{Kanti:2005ja, Harmark:2007jy, Crispino:2013pya}. However, \eqref{Eq:TransmissionCoefficientSquared} is more general and gives the greybody factor at small but non-zero frequencies.

At this stage, a comparison is in order between our results and previous findings in the literature. The results reported in Ref.~\cite{Kanti:2005ja} for the greybody factor are only comparable to ours in the strict zero-frequency limit.
An expression with $\epsilon$-dependent corrections, first found in Ref.~\cite{Harmark:2007jy}, in our notation reads as $\Gamma(\omega)=4(\alpha^2+\epsilon^2)$\,. This agrees with our result \eqref{Eq:TransmissionCoefficientSquared} expanded to second order in both $\alpha$ and $\epsilon$\,.
Similarly, the greybody factor appearing in Ref.~\cite{Crispino:2013pya}, specialized to the case of a minimally coupled scalar,
matches ours to quadratic order in $\epsilon$ and $\alpha$\,, with departures at higher orders due to the fact that their expression, unlike \eqref{Eq:TransmissionCoefficientSquared}, is not symmetric under $\epsilon\leftrightarrow\alpha$\,.
On the other hand, exact flux conservation, $|{\cal R}_\omega|^2+|{\cal T}_\omega|^2=1$\,, obeyed by our transmission and reflection coefficients \eqref{Eq:ScatteringCoefficients}, has not been previously explicitly confirmed.

The non-analyticity of the scattering coefficients in the $\alpha\to0$ limit has non-trivial physical consequences.
The complex reflection coefficient can be re-expressed as ${\cal R}_{\omega}=-|{\cal R}_{\omega}| e^{2i\delta_0}$\,, where $\delta_0$ represents the (frequency-dependent) phase-shift of the spherical wave. Using Eq.~\eqref{Eq:ReflectionCoefficient}, we obtain to leading-order in $\alpha$
\be\label{Eq:phase_shift}
\delta_0\approx\frac{\epsilon}{2\alpha}\ln(2\alpha)~.
\ee
Note that this effect originates from the de Sitter asymptotics, since the phase shift is exactly vanishing for s-wave scattering on a Schwarzschild background. The Wigner time delay can be computed as~\cite{Wigner:1955zz}
\be\label{Eq:TimeDelayRegime1}
\tau\equiv2\frac{\pa \delta_0}{\pa\omega}=2r_h\frac{\pa \delta_0}{\pa\epsilon}\approx r_c\ln(\frac{2r_h}{r_c})~,
\ee
where we used Eq.~\eqref{Eq:phase_shift} in the last step. The effect is more pronounced the smaller the black hole compared to the cosmological horizon.

\section{Large black holes}\label{Sec:regime2}
In this regime, we assume $\epsilon\ll1$\, while both $\alpha$ and $(1-\alpha)$ remain finite.
Notice that, because the cosmological and black-hole horizons are of comparable size, the low-frequency modes we are considering in this section are in the super-horizon regime $\omega\,r_c\ll1$\,.
We partition the spacetime between the two horizons, $r_h$ and $r_c$\,, into three regions:
\begin{itemize}
\item \textit{near region}: $x\ll1$~,
\item \textit{static region}: $\epsilon\ll x\ll1/\epsilon$~,
\item \textit{cosmological region}: $x\gg1$~.
\end{itemize}
The static region overlaps with both near and cosmological regions. Specifically, given that $\epsilon\ll 1$\,, the near-static overlap region is $\epsilon\ll x\ll 1$ and the static-cosmological overlap region is $1\ll x\ll 1/\epsilon$\,.

\subsection{ \texorpdfstring{Near region: $x\ll1$}{Near region}}

In this region, which is located proximally to the black-hole horizon, Eq.~\eqref{Eq:KG_xs_coord} can be approximated as follows
\be\label{Eq:KG_psi_near}
    x^2 \psi_\omega ''(x)-x \psi_\omega '(x)+\left(1+\eta^2 \right)\psi_\omega (x)=0\,,
\ee
where $\eta \equiv \frac{(1 + \alpha + \alpha^2)\epsilon}{(1 - \alpha)(1+2\alpha)}\ll 1$\,.
The general exact solution of this equation is given by
\be\label{Eq:psi_near_soln}
\psi_{\omega}^{\rm(near)}(x)=\tilde{C}_1^{\rm(near)} x^{1-i\eta }+\tilde{C}_2^{\rm(near)} x^{1+i\eta}~,
\ee
with $\tilde{C}_{1,2}^{\rm (near)}$ integration constants. 
Notice that, while we will perform matched asymptotic expansions to leading order only, we have kept the sub-leading $\eta^2$ term in the coefficient of $\psi_\omega$ in Eq.~\eqref{Eq:KG_psi_near}. The reason is that this term, as in the analogous case in Sec.~\ref{Sec:regime1_near}, is necessary for regularity at the horizon: indeed the $x^{1\mp i\eta }$ solutions in \eqref{Eq:psi_near_soln} precisely agree with the regular solutions which behave like $\phi_\omega\sim e^{\mp i\omega r_*}$ at the black-hole horizon.

Away from the horizon, for large $x$, the near-region solution \eqref{Eq:psi_near_soln} exhibits the following asymptotic behaviour,
\be\label{Eq:psi_near_asympt}
    \psi_{\omega}^{\rm(near)}(x) \approx
\left(\tilde{C}_{1}^{\rm(near)}+\tilde{C}_{2}^{(near)}\right)x-i\eta \left(\tilde{C}_{1}^{\rm(near)}-\tilde{C}_{2}^{\rm(near)}\right)x\ln{x}~, 
\ee
provided that $\eta|\ln{x}|\ll1$\,. 
As a result, Eq.~\eqref{Eq:psi_near_asympt} approximates the exact solution to the full dynamics in Eq.~\eqref{Eq:KG_xs_coord} whenever $e^{-1/\eta}\ll x\ll 1$\,, which includes the near-static overlap region.

\subsection{ \texorpdfstring{Static region: $\epsilon\ll x\ll 1/\epsilon$}{Static region}} 

In this region, we can neglect the $\epsilon$-dependent terms in Eq.~\eqref{Eq:KG_xs_coord},
obtaining the following approximation,
\be\label{Eq:KG_psi_static}
\begin{split}
x^2 (1 + x) \, \psi_\omega''(x)
-\frac{1+2\alpha+3\alpha(1+2\alpha)x+\alpha^2 (7+2\alpha)x^2+{\alpha}^2(2+\alpha)x^3}{(1 + \alpha x)\left[1 + 2\alpha  + \alpha (2 + \alpha)x \right]}\,x\, \psi_\omega'(x)&\\[5pt]
+\frac{1+2\alpha+3\alpha(1+2\alpha)x+\alpha^2 (7+2\alpha)x^2+{\alpha}^2(2+\alpha)x^3}{(1 + \alpha x)\left[1 + 2\alpha  + \alpha (2 + \alpha)x \right]}\psi_\omega(x)& =0~,
\end{split}
\ee
whose exact solution reads
\be\label{}
    \psi^{\rm(stat)}_{\omega}(x)=\tilde{C}_{1}^{\rm(stat)} x+\tilde{C}_{2}^{\rm(stat)} x\left[\ln x-\frac{(1-\alpha)(1+2\alpha)}{1+\alpha}\ln (1+x)+\frac{\alpha(1-\alpha )}{(1+\alpha) (2+\alpha)}\ln \big(1+2 \alpha +\alpha (2+\alpha )x\big) \right]\,,
\ee
with integration constants $\tilde{C}_{1,2}^{\rm(stat)}$. This solution has the following asymptotics, respectively, for small and large $x$\,:
\begin{subequations}\label{Eq:Regime2_Static_Asymptotics}
\begin{empheq}[left={\psi_{\omega}^{\rm(stat)}(x)\approx\empheqlbrace}]{align}
\displaystyle
&\,\left[\tilde{C}_{1}^{\rm(stat)}+\dfrac{\alpha(1-\alpha)\ln(1+2\alpha)}{(1+\alpha)(1+2\alpha)}\tilde{C}_{2}^{\rm(stat)}\right]x
+\tilde{C}_{2}^{\rm(stat)}x\ln{x}\,,&\text{as } x\to 0~,\enspace
\label{Eq:Regime2_Static_Asymptotics_a} \\[10pt]
\displaystyle
&\,\left[\tilde{C}_{1}^{\rm(stat)}+\dfrac{\alpha(1-\alpha)\ln\!\big(\alpha(2+\alpha)\big)}{(1+\alpha)(2+\alpha)}\tilde{C}_{2}^{\rm(stat)}\right]x
+\tilde{C}_{2}^{\rm(stat)}\dfrac{\alpha(1+2\alpha)}{(2+\alpha)}x\ln{x}\,,
\quad &\text{as } x\to\infty~.
\label{Eq:Regime2_Static_Asymptotics_b}
\end{empheq}
\end{subequations}
As a result, Eq.~\eqref{Eq:Regime2_Static_Asymptotics_a} approximates the exact solution to the full dynamics in Eq.~\eqref{Eq:KG_xs_coord} whenever $\epsilon\ll x\ll 1$\,, while Eq.~\eqref{Eq:Regime2_Static_Asymptotics_b} does so for $1\ll x\ll 1/\epsilon$\,.

\subsection{\texorpdfstring{Cosmological region: $x\gg 1$}{Cosmological region}}
In this region, which lies proximally to the cosmological horizon, 
Eq.~\eqref{Eq:KG_xs_coord} can be approximated as follows
\be\label{Eq:psifar}
   x^2\psi_{\omega}''(x)-x \psi_{\omega}'(x)+\left(1+\tilde{\eta}^2\right) \psi_{\omega}(x)=0~,
\ee
where $\tilde{\eta}\equiv \frac{(1+\alpha+\alpha^2)\epsilon}{\alpha(1-\alpha)(2+\alpha)}$~.
Notice that while $\tilde{\eta}\ll1$ we have nevertheless kept the sub-leading $\tilde{\eta}^2$ term in the coefficient of $\psi_\omega$ in Eq.~\eqref{Eq:psifar} for the same reason that we did similarly in the derivation of the near-region equation as well. Namely, the exact solution to Eq.~\eqref{Eq:psifar} is then given by
\be\label{Eq:KG_psi_cosmo}
    \psi_{\omega}^{\rm(cosmo)}(x)=\tilde{C}_{1}^{\rm(cosmo)}x^{1-i\tilde{\eta}}+\tilde{C}_{2}^{\rm(cosmo)}x^{1+i\tilde{\eta}}~,
\ee
with $\tilde{C}_{1,2}^{\rm(cosmo)}$ integration constants, and this is in agreement with the regularity condition $\phi\sim e^{\mp i\omega r_*}$ at the cosmological horizon.

Away from the horizon, for small $x$, the cosmological-region solution \eqref{Eq:KG_psi_cosmo} exhibits the following asymptotic behaviour,
\be\label{Eq:inner_cosmo}
    \psi_{\omega}^{\rm(cosmo)}(x) \approx \left(\tilde{C}_{1}^{\rm(cosmo)}+\tilde{C}_{2}^{\rm(cosmo)}\right)x-i\tilde{\eta} \left(\tilde{C}_{1}^{\rm(cosmo)}-\tilde{C}_{2}^{\rm(cosmo)}\right)x\ln {x}~,
\ee
provided that $\tilde{\eta}|\ln x|\ll1$\,.
As a result, Eq.~\eqref{Eq:inner_cosmo} approximates the exact solution to the full dynamics in Eq.~\eqref{Eq:KG_xs_coord} whenever $1\ll x\ll e^{1/\tilde{\eta}}$\,, which includes the static-cosmological overlap region.

\subsection{Matching formulae and scattering coefficients}

\begin{figure}
    \centering
\subfloat[]{\scalebox{0.555}{\includegraphics{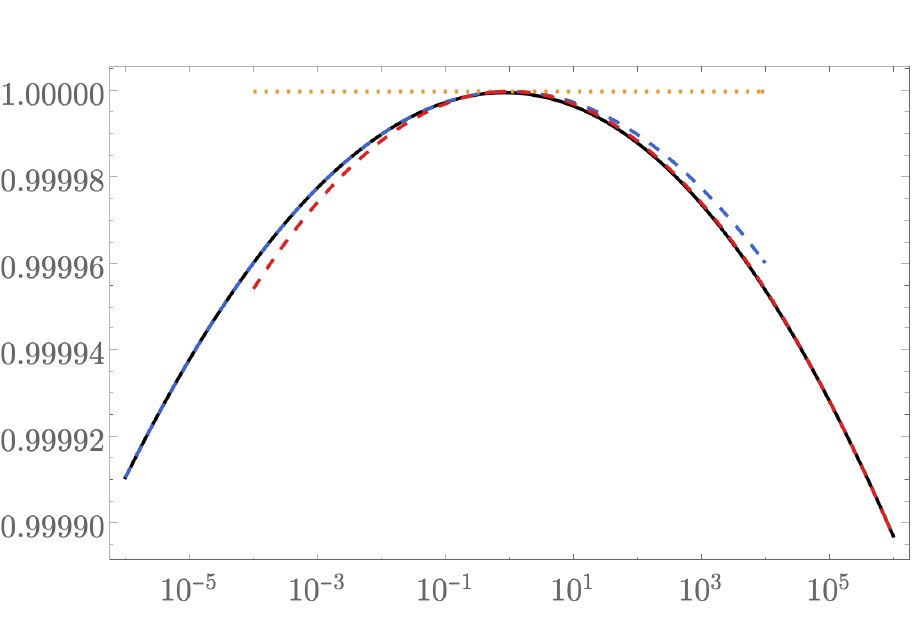}}}
\qquad
\subfloat[]{\scalebox{0.54}{\includegraphics{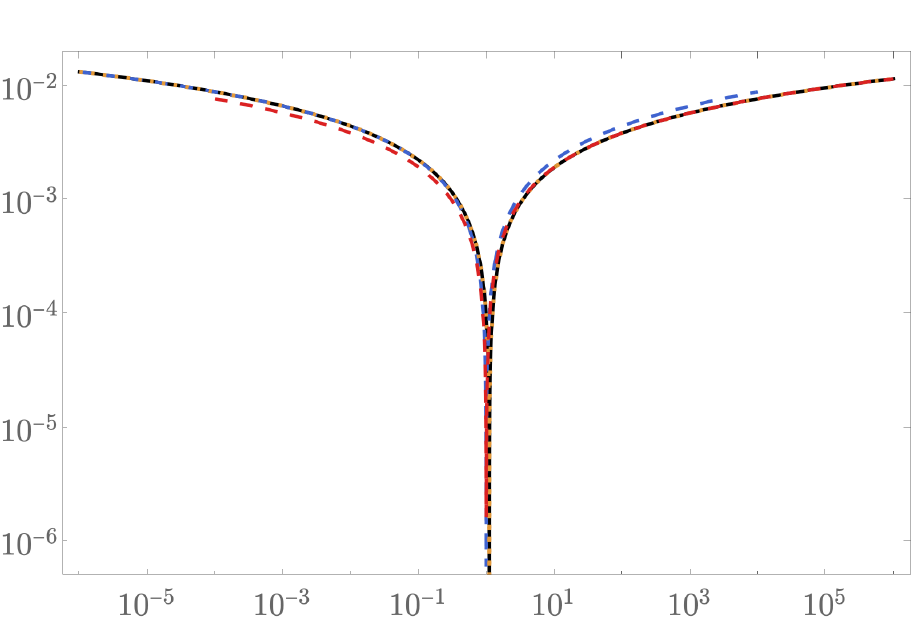}}}
\caption{The plots show the agreement, in the large black-hole regime, between a numerical solution for $\phi(x)$ and our analytical approximations obtained with matched asymptotic expansions, for parameters $(\epsilon,\alpha)=(10^{-4},0.9)$\,. The left and right panels, respectively, show the real and imaginary parts of $\phi$ as a function of $x$\,. The thick black curve represents the numerical solution, the dashed blue curve the near-region approximation, the dot-dashed orange curve the static-region approximation, while the dashed red curve corresponds to the cosmological-region approximation.}\label{Fig:Regime2}
\end{figure}

In the the near-static overlap region $\epsilon\ll x\ll1$\,, matching the solutions~\eqref{Eq:psi_near_asympt} and~\eqref{Eq:Regime2_Static_Asymptotics_a} we obtain
\be\label{Eq:MatchingConditions1_Regime2}
\begin{split}
\tilde{C}_{1}^{\rm(near)}+\tilde{C}_{2}^{\rm(near)}&=\tilde{C}_{1}^{\rm(stat)}+ \dfrac{\alpha(1-\alpha)  \ln (1+2 \alpha)}{(1+\alpha) (2+\alpha)}\tilde{C}_{2}^{\rm(stat)}~,\\
\qquad -\left(\tilde{C}_{1}^{\rm(near)}-\tilde{C}_{2}^{\rm(near)}\right)i\eta&=\tilde{C}_{2}^{\rm(stat)} ~,
\end{split}
\ee
while in the static-cosmological overalp region $1\ll x\ll 1/\epsilon$, matching the solutions~\eqref{Eq:Regime2_Static_Asymptotics_b} and~\eqref{Eq:inner_cosmo} we obtain
\be\label{Eq:MatchingConditions2_Regime2}
\begin{split}
\tilde{C}_{1}^{\rm(stat)}+\dfrac{\alpha (1-\alpha )  \ln [\alpha  (2+\alpha)]}{(1+\alpha) (2+\alpha)}\tilde{C}_{2}^{\rm(stat)}  &= \tilde{C}_{1}^{\rm(cosmo)}+\tilde{C}_{2}^{\rm(cosmo)}~,\\
\dfrac{\alpha(1+2\alpha)   }{2+\alpha } \tilde{C}_{2}^{(stat)} &= -  \left(\tilde{C}_{1}^{(cosmo)}-\tilde{C}_{2}^{(cosmo)}\right) i\tilde{\eta}~.
\end{split}
\ee
Eliminating from the above the amplitudes of the static solution yields the relation of amplitudes at the black-hole and cosmological horizons.
For example, imposing purely ingoing boundary conditions at the black-hole horizon, $\tilde{C}_2^{\rm(near)}=0$\,, the solution for the amplitudes of the transmitted and reflected waves in terms of the incident one is given by
\begin{align}
\tilde{C}_1^{\rm(near)}&=\frac{ 2(1+\alpha) (2+\alpha) (1+2 \alpha)}{(1+\alpha)(2+\alpha) (1+2\alpha) \left(1+\alpha ^2\right)+i \alpha  \left(1+\alpha+\alpha^2\right) \epsilon  \ln \left(\frac{1+2 \alpha}{\alpha(2+\alpha)}\right)}\tilde{C}_1^{\rm(cosmo)} ~,\\
\tilde{C}_2^{\rm(cosmo)}&=\frac{(1-\alpha) (1+\alpha)^2(2+\alpha) (1+2 \alpha)+i \alpha  \left(1+\alpha+\alpha ^2\right) \epsilon  \ln \left(\frac{1+2\alpha}{\alpha(2+\alpha)}\right)}{(1+\alpha) (2+\alpha) (1+2\alpha) \left(1+\alpha^2\right)+i \alpha  \left(1+\alpha+\alpha ^2\right)\epsilon  \ln \left(\frac{1+2 \alpha}{\alpha(2+\alpha)}\right)}\tilde{C}_1^{\rm(cosmo)}  ~.
\end{align}
In Fig.~\ref{Fig:Regime2} we show the agreement between our analytical solution obtained region-wise using the above connection formulae and a numerical solution of Eq.~\eqref{Eq:KG_xs_coord}. 
We note in passing that the static region may actually be extended beyond its definition above, $\epsilon\ll x\ll1/\epsilon$\,, to the much broader region $e^{-1/\eta}\ll x\ll e^{1/\tilde{\eta}}$\,.

The transmission and reflection coefficients are defined by
\be\label{Eq:ScatteringCoefficients2}
{\cal T}_\omega \equiv \frac{r_h \tilde{C}_1^{\rm(near)}}{r_c \tilde{C}_1^{\rm(cosmo)}}~,\quad
{\cal R}_\omega \equiv \frac{r_c \tilde{C}_2^{\rm(cosmo)}}{r_c \tilde{C}_1^{\rm(cosmo)}}~.
\ee
Hence we obtain
\begin{subequations}
\begin{align}
|{\cal T}_\omega|^2&=\frac{4 \alpha ^2 (1+\alpha)^2 (2+\alpha)^2 (1+2 \alpha)^2}{(1+\alpha)^2 (2+\alpha )^2 (1+2 \alpha)^2 \left(1+\alpha ^2\right)^2+\alpha ^2 \left(1+\alpha+\alpha ^2\right)^2 \epsilon ^2 \ln ^2\left(\frac{1+2 \alpha}{\alpha  (2+\alpha)}\right)}~,\label{Eq:transmission1}\\
|{\cal R}_\omega|^2&=\frac{(1-\alpha)^2(1+\alpha)^4(2+\alpha)^2(1+2\alpha)^2+\alpha ^2 \left(1+\alpha+\alpha ^2\right)^2 \epsilon ^2 \ln ^2\left(\frac{1+2 \alpha}{\alpha  (2+\alpha)}\right)}{(1+\alpha)^2 (2+\alpha )^2 (1+2 \alpha)^2 \left(1+\alpha ^2\right)^2+\alpha ^2 \left(1+\alpha+\alpha ^2\right)^2 \epsilon ^2 \ln ^2\left(\frac{1+2 \alpha}{\alpha  (2+\alpha)}\right)}~.\label{Eq:reflection1}
\end{align}    
\end{subequations}
We note that flux conservation $\lvert{\cal T}_\omega\lvert^2+\lvert{\cal R}_\omega\lvert^2=1$ is satisfied exactly.
The greybody factor is $\Gamma(\omega)=|{\cal  T}_\omega|^2$\,. Thus we find that, to leading order in the low-frequency limit $\epsilon\to 0$\,, the greybody factor is
\be\label{Eq:LowFrequencyLimit2}
\Gamma(\omega)\approx \frac{4\alpha^2}{(1+\alpha^2)^2}+{\cal O}(\epsilon^2)~.
\ee
The expression \eqref{Eq:LowFrequencyLimit2} coincides with the zero-frequency limit of the greybody factor obtained in  Eq.~\eqref{Eq:LowFrequencyLimit1} for small black holes, thus showing the robustness of this result, which holds for black holes of all sizes relative to the cosmological horizon.

We can compute the Wigner time delay in this regime following similar steps as in Section~\ref{Sec:Regime1_Matching}. To leading order in $\epsilon$\,, we obtain
\be
\tau\approx  r_c\frac{2\alpha^4(1+\alpha+\alpha^2)\,\ln{\left(\frac{1+2\alpha}{\alpha(2+\alpha)}\right)}}{(1-\alpha)(1+\alpha)^2(2+\alpha)(1+2\alpha)(1+\alpha^2)}~.
\ee
For $\alpha\approx1$\,, this boils down to $\tau\approx r_c\left(\frac{1}{18}-\frac{1}{12}(1-\alpha)\right)$\,. Thus, unlike the small black-hole regime, where $\tau$ diverges as the size of the black hole shrinks to zero (relative to the cosmological horizon), in this case the limit is finite and proportional to $r_c$\,.

\section{Conclusions}\label{Sec:conclusions}
We performed an analytical study of the classical scattering of a massless scalar field off a Schwarzschild-de Sitter black hole in the low-frequency regime using the method of matched asymptotic expansions. 
This represents a step forward in the analytic control of the dynamics for this system, and sheds light on the interplay between cosmological and near-horizon dynamics of black holes.
Our focus is on radial scalar perturbations (i.e., zero angular momentum), as the transmission of higher multipoles through the effective potential barrier is strongly suppressed in the low-frequency limit. The dynamics is governed by two dimensionless parameters, $\epsilon=\omega\, r_h$ and $\alpha=r_h/r_c$\,. We identified two regimes of interest, corresponding to small and large black holes, with $\alpha\ll1$ and $\alpha\sim{\cal O}(1)$\,, respectively. In each regime we calculate analytically the scattering coefficients and greybody factor as functions of $\epsilon$ and $\alpha$\,.
The presence of a positive cosmological constant has a nontrivial effect on the evolution of the system, even in the zero-frequency limit, through $\alpha\neq0$\,. In particular, for small black holes, the reflection coefficient has an essential singularity in the $\alpha\to0$ limit, which in turn is responsible for a non-trivial time delay of the reflected wave. This effect is completely absent in a Schwarzschild geometry.

The method of matched asymptotic expansions gives a good approximation to the solution of the dynamics, provided that the various asymptotic regions have non-vanishing overlaps and that their respective asymptotic solutions are smoothly matched in these overlap regions. The method is apt to handle singular perturbation problems and has been applied to the dynamics of matter fields and gravitational waves in black-hole spacetimes as early as ~\cite{Starobinskii:1973vzb, Starobinskil:1974nkd, Teukolsky:1974yv, Unruh:1976fm}.
Our rigorous application of the method offers a significant advantage compared to other methods based on an integral representation of the solutions to the scalar wave equation \cite{Anderson:2015bza}, both in terms of simplicity of the functional form of the solutions (and associated physical observables), and in terms of analytic control of the system at small but non-zero frequency.

At $\epsilon=0$\,, Eq.~\eqref{Eq:LowFrequencyLimit1} confirms results previously obtained, with less rigorous methods, in Refs.~\cite{Kanti:2005ja,Harmark:2007jy} for the greybody factor of small black holes. 
In the case of small black holes, our implementation of matched asymptotic expansions enjoys several advantages over previous treatments. We are using two manifestly overlapping `near' and `far' regions, whereas, e.g., in~\cite{Harmark:2007jy} two of the solutions matched are derived in two non-overlapping regions.
Also, our `far' solution is sensitive to both characteristic scales of the problem and interpolates between the cosmological asymptotics in the proximity of the de Sitter horizon and an intermediate asymptotics between the de Sitter and black-hole horizons, where it is smoothly joined to the `near' asymptotics. 

For large black holes, with $\alpha$ and $(1-\alpha)$ $\sim{\cal O}(1)$\,, low-frequency modes are super-horizon modes. In this case, which has not been considered previously, we show that the zero-frequency limit of the greybody factor, Eq.~\eqref{Eq:LowFrequencyLimit2}, is
the same as for small black holes.
Super-horizon modes of near-extremal black holes with $(1-\alpha)\ll1$ are also included, so long as the condition $\epsilon\ll(1-\alpha)$ is satisfied. However, a full study of the dynamics of perturbative modes on a near-extremal Schwarzschild-de Sitter black hole (and the corresponding $dS_2\times S^2$ geometry) requires separate treatment, analogous to the one for near-extremal Reissner-Nordstr{\"o}m black holes in Refs.~\cite{Porfyriadis:2018yag,Porfyriadis:2018jlw}.

A natural extension of the present work is to compute the impact of the scattered scalar on the black hole geometry, along the lines of
\cite{deCesare:2023rmg,deCesare:2024csp}, and identify appropriate physical observables associated with the backreaction.
The generalization of our analysis to a massive scalar field, which will be the subject of future work, is also of interest since light scalars may play the role of dark matter and form long-lived clouds around black holes, with non-trivial effects on their dynamics and on the gravitational-wave signal emitted by black-hole binaries \cite{Clough:2019jpm,Traykova:2023qyv,Aurrekoetxea:2023jwk,Blas:2024duy}.

\section*{Acknowledgments}
MdC acknowledges support from INFN iniziativa specifica GeoSymQFT. This research is implemented in the framework of H.F.R.I. call “Basic
research Financing (Horizontal support of all Sciences)” under the National
Recovery and Resilience Plan ``Greece 2.0'' funded by the European Union
---NextGenerationEU (H.F.R.I. Project Number: 15384). APP is also partially
supported by UoC grant number 12030.

\bibliographystyle{apsrev4-1}
\bibliography{bibliography}

\end{document}